\documentstyle[pra,multicol,aps]{revtex}
\begin{document}
\input{epsf.tex}
\epsfverbosetrue
\title{Solitonic Gluons}
\author{Elena A. Ostrovskaya and Yuri S. Kivshar}
\address{Optical Sciences Centre, Research School of
Physical Sciences and Engineering\\
Australian National University, Canberra
ACT 0200, Australia}
\author{Zhigang Chen$\dagger$ and Mordechai Segev$\ddagger$}
\address{Department of Electrical Engineering and Center for
Photonics and Optoelectronic Materials\\ Princeton University,
Princeton, New Jersey 08544}
\maketitle
\begin{abstract}
We describe a physical mechanism for creating multi-soliton bound states
where  solitons are glued together by attraction between the beams they
guide, {\em  solitonic gluons}. We verify the concept of the solitonic
gluons experimentally,  observing a suppression of the repulsion between
dark solitons due to an  attractive force between bright guided beams.
\end{abstract}
\vspace{-1.0cm}
\begin{multicols}{2}
\narrowtext
It has been  established for many models of nonlinear physics that solitons
behave as  effective particles when interacting with external fields or
other solitons  \cite{kiv_mal}. In the case of bright solitons described by
the  nonlinear Schr{\"o}dinger (NLS) equation, the coherent interaction of
solitons depends on the  relative phase between them, so that identical
solitons with  opposite phases repel each other, whereas in-phase solitons
attract each other \cite{chu}. Interaction of dark solitons is
unconditionally  repulsive, in all types of models described by the
generalized NLS equation  \cite{dark_review}.

If a nonlinear system supports propagation of two, or more, waves of
different  frequencies or polarization, vector solitons consisting of more
than one  components can be formed.  Indeed, when two vector solitons are
closely  separated, they may form a bound state if the sum of all forces
acting between  different soliton components is zero. As an example, let us consider
interaction of two vector solitons consisting of dark and bright
components in a defocusing optical medium. Two dark solitons always repel
each  other, and as a result, they can not form a bound state
\cite{dark_review}.  However, if we introduce out-of-phase bright components guided
by each of the dark solitons, their attractive interaction creates a proper
balance of forces, which results in a stationary two-soliton bound state (see Fig. 1).

The mechanism leading to the formation of bound states of the compound
solitons  described above, makes it somewhat tempting to draw an analogy
between the soliton interaction and the fundamental theory of quarks and
gluons  -- quantum chromodynamics. As is well known \cite{gluons}, gluons
mediate
the strong force and are responsible for binding the quarks into protons
and  neutrons, for example. Unless forming a bound state with quarks,
gluons do not  exist as separate particles.  Taking these properties of
gluons into consideration, we believe that it would
be plausible to call bright components guided by dark solitons in a
defocusing  medium `solitonic gluons'. On the one hand, they do not `live'
in a defocusing  medium by themselves, i.e. they diffract when separated
from the dark solitons;  and on the other hand, they `glue' the dark
solitons together in a bound state  which otherwise can not be formed.
\vspace{-0.3cm}
\begin{figure}
\setlength{\epsfxsize}{4.0cm}
\centerline{\epsfbox{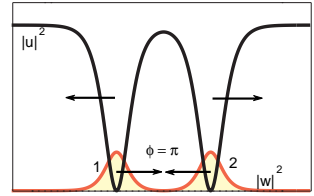}}
\vspace{0.2cm}
\caption{Schematic presentation of the force balance for the interacting
dark-bright compound solitons. Shown are the intensities of dark and bright
components in the fields $u$ and $w$, respectively; $\phi$ is a relative
phase  between the bright guided beams 1 and 2.}
\label{fig1}
\end{figure}
\vspace{-0.4cm}
To demonstrate this general concept on a particular example and verify it
later  experimentally, we consider two incoherently interacting linearly
polarized  beams in a photorefractive medium.  The corresponding physical
model is  described by the equations for the normalized envelopes $U$ and
$W$ \cite{segev}:
 \begin{equation}
\label{eq_UW}
\begin{array}{l} {\displaystyle
i \frac{\partial U}{\partial z} + \frac{1}{2} \frac{\partial^2
U}{\partial x^2} + \frac{ \beta (1 + \rho) U}{ 1 + |U|^2 +
|W|^2}  \hspace{2.4mm} =  0,} \\*[9pt] {\displaystyle i
\frac{\partial W}{\partial z} + \frac{1}{2} \frac{\partial^2
W}{\partial x^2} + \frac{ \beta (1 + \rho) W}{ 1 + |U|^2 +
|W|^2}  \hspace{2.4mm} = 0.}
\end{array}
\end{equation}
Here $\rho = I_{\infty}/I_d$, $\beta = (k_0x_0)^2 n_e^4 r_{33} E_0/2$, where
$I_{\infty}$ stands for the total power density in the limit $x \rightarrow
\infty$, $I_d$ is the so-called dark irradiance, $k_0$ is the propagation
constant, $x_0$ is the spatial width of the beam, and $n_e^4 r_{33}E_0$ is a
correction to the refractive index due to the external field applied to a
crystal along the $x$-axis \cite{segev}.

Solitonic solutions of Eqs. (\ref{eq_UW}) can be sought for in the form $U=
\sqrt{s} \, u \, \exp [i\sigma (1-s) z]$ and $W = \sqrt{s} \, w \,  \exp
[i\sigma (1-s\lambda) z]$, where $s<1$ and $\lambda<1$ are two dimensionless
parameters and $\sigma \equiv {\rm sgn} \beta$ defines the type of
nonlinearity,
namely defocusing ($\sigma=+1$) or focusing ($\sigma=-1$). Amplitudes $u$ and
$w$ satisfy the normalized equations:
\begin{equation}
\label{eq_uw}
\begin{array}{l} {\displaystyle
i \frac{\partial u}{\partial z} + \frac{1}{2} \frac{\partial^2 u}{\partial
x^2}
- \frac{ \sigma (|u|^2 + |w|^2)u}{ 1 + s(|u|^2 +|w|^2)} + u \hspace{2.4mm} =
0,} \\*[9pt] {\displaystyle i \frac{\partial w}{\partial z} + \frac{1}{2}
\frac{\partial^2 w}{\partial x^2} - \frac{\sigma (|u|^2 + |w|^2)w}{ 1 +
s(|u|^2
+ |w|^2)} + \lambda w \hspace{2.4mm} = 0,}
\end{array}
\end{equation}
where the spatial coordinates $z$ and $x$ are measured in the units of
$s/|\beta| (1 + \rho)$ and $\{s/[|\beta|(1+\rho)]\}^{1/2}$, respectively.
In Eq.
(\ref{eq_uw}), the parameter $s$ characterizes nonlinearity saturation.

First, assuming $\sigma = +1$, we look for stationary, $z$--independent solutions of the system
(\ref{eq_uw}) in the form of dark-bright solitons,  where
$u(x)$  has nonvanishing  asymptotics but  $w(x)$ vanishes for $|x|
\rightarrow \infty$.    Then, a compound dark-bright solitary wave $(u,w)$,
emerges from the  one\--component dark solution $(u,0)$ at a certain
bifurcation point $\lambda =
\lambda_0(s)$.  Near the bifurcation point, we can treat the dark component
as an  effective waveguide which guides the spatially localized mode $w$.
Such compound  solitons form a continuous family in $\lambda$, for a fixed
$s$. The existence
of this type of vector dark-bright solitons in the photorefractive media
has  been earlier established theoretically \cite{segev} and verified
experimentally  \cite{bright_dark}. The families of two-component
dark-bright solitons have been  numerically  found
in Ref. \cite{ol_ost}.

Interaction between two weakly overlapping dark-bright solitons can be
described by the effective interaction energy as a function of the soliton
separation  $x_0$,
$V_{\rm eff}(x_0) = V_d(x_0) + a^2 V_b(x_0) \cos \phi$,
where the first and second terms stand for the interaction between dark and
bright soliton components, respectively, with $\phi$ being a relative phase
between the bright components of the equal amplitude $a$. In the  case of
small $s$, the effective interaction energies $V_d(x_0)$ and $V_d(x_0)$
can be found analytically \cite{ol_ost}, $V_d(x_0) \approx 2 \exp
(-4x_0)$ and $V_b(x_0) \approx 4 x_0 \exp (-2 x_0)$. Thus, interaction
between two dark-bright solitons is crucially
affected by the forces acting between the bright components guided by the
closely separated dark solitons. If the bright components are out of phase
($\phi =\pi$), an attractive force acting between them can balance (or even
overbalance) the repulsion between the dark solitons, thus creating a bound
state. This observation has been confirmed for Eqs. (\ref{eq_uw}) by
numerically identifying the families of two-soliton bound states
\cite{ol_ost}.

To demonstrate the concept of force balance, we have investigated
numerically the interaction of a pair of dark  solitons generated by a
localized input in the shape of a box-like notch on a cw  background.
Figures 2(a) and 2(b) show the results of our simulations  for two distinct
cases. First, the box-like input is launched without a bright  component.
It generates a pair of scalar (one-component) dark solitons which  move
away from eachother as they propagate [see Fig. 2(a)] due to their mutual
repulsion \cite{dark_review}. However, the propagation dynamics is
drastically  modified by introducing an additional bright beam. Figure 2(b)
shows  the results for the input field as in Fig. 2(a) but with the added
$w-$component  in the form of the first (one-node) mode of the symmetric
box-shaped waveguide.  Such an amplitude profile of the bright component
has been chosen to mimic a
solitonic gluon binding two closely separated dark solitons. As is seen
from  Fig. 2(b), the attraction between out-of-phase bright guided beams
compensates  for the repulsive force between dark solitons, so that  two
dark-bright solitons propagate almost in  parallel.
\vspace{-0.2cm}
\begin{figure}
\vspace{-0.2cm}
\setlength{\epsfxsize}{6.7cm}
\centerline{\epsfbox{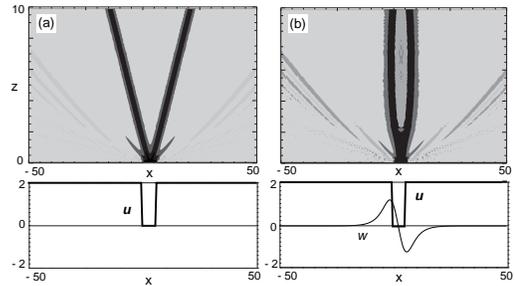}}
\caption{ Generation of a pair of dark solitons from  a localized box-like
input: (a) no bright component introduced, (b) in the  presence of the
bright component. Bottom row: corresponding input beam profiles.}
\label{fig2}
\vspace{-0.2cm}  
\end{figure}
\vspace{-0.4cm}
\begin{figure}
\setlength{\epsfxsize}{6.7cm}
\centerline{\epsfbox{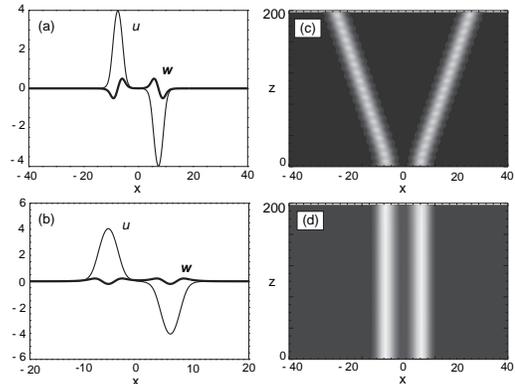}}
\caption{Bound states of bright solitons in the model (\ref{eq_uw}) at
$\sigma =  - 1$. (a),(b) Examples of compound solitons with the first and
second guided  modes, respectively. (c),(d) Interaction of two bright
solitons from (a),  without and with the gluon component, respectively.}
\label{fig3}        
 \end{figure}
\vspace{-0.4cm}
The concept of solitonic gluons can be readily extended to describe
multi-soliton bound states composed of various types of bright vector
solitons \cite{snyder,yang}.  Bright solitons of Eqs. (\ref{eq_uw})  are
possible for a self-focusing nonlinearity,
$\sigma=-1$. Continuous in $\lambda$, families of such solutions originate
at the points of bifurcation of the scalar
($u-$component only) bright soliton families. Similar to the case of
dark-bright  solitons, the $w-$ field near the bifurcation points can be
described as a  linear mode of an effective waveguide induced by the
$u-$soliton, the  latter can  support more than one mode, depending on the
value of $s$.

A rich variety of the soliton structures, consisting of a fundamental
bright  soliton and a $n$--node mode of the soliton-induced waveguide,
results in complex  shapes of solitonic gluons of the  focusing
nonlinearity. Interaction between the bright  components depends on their
relative phase, and therefore the bound states of
bright vector solitons should be constructed from out-of-phase (mutually
repelling) and  in-phase (mutually attracting) components. Some vector
solitary waves of this  type have already been described  \cite{yang}. In
Figs. 3(a, b) we present two examples of such soliton bound  states, found
numerically for Eqs. (\ref{eq_uw}), where the repulsion of the
out-of-phase fundamental (no nodes) solitons of the $u-$ component is
balanced  by an attractive force acting between two types of solitonic
gluons, formed by  the in-phase first or second modes of effective
soliton-induced waveguides.

Numerical simulations of the beam propagation, with the input of the type
presented in Fig. 3(a), revealed  a drastic change in the soliton
interaction as shown in Figs. 3(c,d) [cf. Fig. 2(a,b)].
The binding force of the solitonic gluons makes the otherwise
mutually repelling out-of-phase solitons propagate in parallel [Fig. 3(d)].

To verify the concept of solitonic gluons experimentally, we have
studied the interaction of dark solitons in a photorefractive medium, using
the
experimental setup for generating dark-bright soliton pairs earlier
reported in
Refs. \cite{bright_dark,zhigang}. A cw argon-ion laser beam  (488 nm) is
collimated and split by a polarizing beam splitter. The ordinary polarized
beam
is used as uniform background illumination to  mimic the dark irradiance, and
the extraordinary polarized beam is split into two soliton-forming beams.
One of
these two beams is used to generate a dark notch on reflection, by
illuminating
a wire. The dark notch is then imaged onto the input face of a SBN:61 crystal
with the background beam covering the entire input face. One half of the other
beam (the bright component) goes through a tilted thin glass that introduces a
$\pi$ phase jump at its centre, thereby creating a field profile resembling
the
first mode of a box-like waveguide, i.e. a bright beam with a node [Fig. 4(a)]. We make
the dark and bright input beams mutually incoherent by having their optical
path
difference greatly exceed the coherence length of the laser.
\vspace{-1.0cm}
\begin{figure}
\setlength{\epsfxsize}{8.0cm}
\centerline{\epsfbox{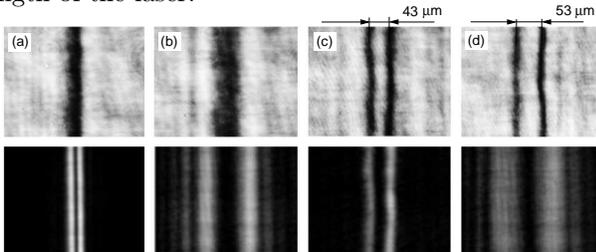}}
\caption{Photographs showing the effect of the solitonic gluons via coupling
of a  bright (lower row) and dark (upper row) incoherently interacting beams:
(a) an  input at $z=0$ mm, and three outputs at $z=11.7$ mm for (b) normal
diffraction of coupled beams, (c) stationary two-soliton bound state, (d) decoupled beam
propagation.}
\label{fig4}
\end{figure}
\vspace{-0.4cm}
Figures 4(a-d) show photographs of the bright and dark beams taken on the
input [Fig. 4(a)] and output [Figs. 4(b,c,d)] faces of a 11.7-mm-long SBN
crystal.
Without a dc field, both beams merely diffract in a similar fashion,
as shown in Fig. 4(b). When we launch only a dark-notch-bearing beam
in the presence of an appropriate dc field in the $x$ direction
($||c$ axis), it generates a Y-junction created by a pair of
repelling dark solitons, as is seen in Fig. 4(d) (upper row) and earlier
reported in Ref. \cite{zhigang}. On the other hand, the uncoupled bright beam
diffracts even more strongly due to effectively defocusing nonlinearity, Fig.
4(d) (lower row).

Repulsion of a pair of dark solitons created by a wire is suppressed
dramatically when we simultaneously launch a bright beam in the form of a
second waveguide mode, thus creating two bright beams with opposite phases.
As  can be seen from  Figs. 4(c) and 4(d), the output distance between the
dark solitons is subsequently reduced from $53 \mu m$ to $43 \mu m$.  In
addition,  we have verified that when the repulsion between the dark
solitons is enhanced when the bright beams are in-phase.

In conclusion, we have suggested and verified experimentally
a concept of solitonic gluons featuring a  balance of attractive and
repulsive forces acting between different components  of compound solitons
as a physical mechanism responsible for the formation of bound states of
vector  solitons.  This novel type of interaction of compound solitons is a
rather fundamental  phenomenon which should also be observed for many other
multi-component nonlinear models, e.g. incoherent solitons.

The work has been supported by the Australian Photonics Cooperative
Research Centre. The authors are indebted to D. Christodoulides, M.
Mitchell, and A. Sheppard for useful discussions.

$\dagger$ Present address: Department of Physics and Astronomy, San
Francisco  State University, San Francisco, CA 94132, USA

$\ddagger$ Present address: Physics Department, Technion-Israel Institute
of  Technology, Haifa, 3200, Israel

\end{multicols}
\end{document}